\documentclass[twocolumn,showkeys,showpacs,prb]{revtex4}
\usepackage{amsfonts, amsmath, latexsym}
\usepackage[dvips]{graphicx}
\usepackage{amssymb}

\newtheorem{1}{Theorem}

\begin{document}

\title{Optimal series representations for numerical path integral simulations}

\author{Cristian Predescu}
\author{J. D. Doll}
 
\affiliation{
Department of Chemistry, Brown University, Providence, Rhode Island 02912
}
\date{\today}

\begin{abstract}
By means of the Ito-Nisio theorem, we introduce and discuss a 
general approach to series representations of path integrals. We then
argue that the optimal basis for both ``primitive'' and partial averaged
approaches is the Wiener sine-Fourier basis. The present analysis also
suggests a new approach to improving the convergence of primitive path
integral methods. Current work indicates that this new technique, the
``reweighted'' method, converges as the cube of the number of path
variables for ``smooth'' potentials. The technique is based on a special
way of approximating the Brownian bridge which enters the Feynman-Ka\c{c}
formula and it does not require the Gaussian transform of the potential
for its implementation.
\end{abstract}

\pacs{02.70.Ss, 05.30.-d}
\keywords{density matrix; path integrals; random series; Monte Carlo}

\maketitle

\section{Introduction} 
\newcommand{\ud}{\mathrm{d}}

Numerical simulations based on the path integral approach have
proved highly successful in the calculation of thermodynamic properties for complex,
many-body quantum systems (see Refs.~\onlinecite{Cep95, Mie01} and the
cited bibliography).  Mainly the result of Feynman\cite{Fey48} and
Ka\c{c},\cite{Kac50} the centerpiece of the theory is the fact that the
density matrix of a monodimensional system can be written as the
expectation value of a suitable functional of a standard Brownian
bridge~$\{B_u^0,\, 0\leq u\leq 1\}$. More precisely, if~$\{B_u,\, u\geq
0\}$ is a standard Brownian motion starting at zero, then the Brownian
bridge is the stochastic process~$\{B_u |\,B_1=0,\, 0 \leq u \leq 1\}$
i.e., a Brownian motion conditioned on~$B_1=0$.\cite{Dur96} In this
paper, we shall reserve the symbol~$\mathbb{E}$ to denote the expected
value (average value) of  a certain random variable against the
underlying probability measure of the Brownian bridge~$B_u^0$. For a
monodimensional canonical ensemble characterized by the inverse
temperature $\beta=1/(k_B T)$ and made up of identical particles of
mass~$m_0$ moving in the potential~$V(x)$, the
Feynman-Ka\c{c} density matrix formula reads:\cite{Fey48,Kac50,Sim79}
\begin{equation}
\label{eq:1}
\frac{\rho(x,x';\beta)}{\rho_{fp}(x,x';\beta)}=\mathbb{E}\exp\left\{-\beta\int_{0}^{1}\! \! 
V\Big[x_{0}(u)+\sqrt{\frac{\beta\hbar^2}{m_0}} B_u^0 \Big]\ud u\right\},
\end{equation}
where $\rho_{fp}(x,x';\beta)$ stands for the density matrix of a similar free particle canonical ensemble, while~$x_{0}(u)$ is a shorthand for $x+(x'-x)u$.

Current research is focused on the development of accurate, 
finite-dimensional 
approximations of the stochastic integrals that appear in Eq.~1 and 
in related thermodynamic
expressions.   The importance of Eq.~1 as given here consists of 
the fact that the Brownian motion, hence the Brownian bridge, are well
understood mathematical objects, which can be simulated by a variety of
means. The discussion in the present paper is based on the random series
technique as a general representation scheme for the Brownian
bridge~$B_u^0$.  The approach is particularly interesting, as it is
directly related to the ``path integral'' concept, and can be justified
by means of the Ito-Nisio theorem,\cite{Kwa92} whose statement is
presented in Appendix~A. 

 We consider a number of questions related to the random series implementation of the 
Feynman-Ka\c{c} formula. The so called primitive\cite{Dol84} and partial
averaging\cite{Dol85}  techniques, developed initially for the Fourier
path integral (FPI) method,\cite{Dol84} are generalized here for
arbitrary series representations. Then, we address the question of
whether or not there exists a preferred basis within which to implement
the two techniques. We present strong evidence  suggesting that the
fastest convergent series for each method is the Wiener series on which
the Fourier path integral  approach is based.   Finally, we introduce a
new, non-averaging technique called the reweighted FPI method  in order
to improve the  convergence of primitive FPI. 
	
Motivated by the optimality of the Wiener series, we undertake 
the task of establishing numerically the asymptotic rate of convergence
for the three FPI methods: the primitive FPI, the partial averaging FPI
(PA-FPI), and the reweighted FPI (RW-FPI).   The asymptotic rate of
convergence of the primitive FPI  was extensively studied\cite{Mie01,
Ele98} and is known to  be $\mathcal{O}(1/n)$ for sufficiently smooth
potentials.  However, there are at present no analytical or numerical
studies concerning  the exact asymptotic behavior of the PA-FPI method.
For the particular case of the harmonic oscillator, it is known that the
asymptotic  rate of convergence is $\mathcal{O}(1/n^3)$. (The reader
should not  mistake the full PA-FPI for the so called gradient
corrected PA-FPI,  which was shown to converge as fast as
$\mathcal{O}(1/n^2)$ in Ref.~\onlinecite{Ele98}  for potentials having
continuous second-order derivatives). To cope with the numerical difficulties encountered, we develop a Monte Carlo technique
which allows us to study  the asymptotic  behavior of the PA-FPI and
RW-FPI methods, at least for single-well  potentials. With its help, we
find strong numerical evidence suggesting that the  asymptotic rate of
convergence for both PA-FPI and RW-FPI approaches is $\mathcal{O}(1/n^3)$
for sufficiently smooth potentials.  To our knowledge, RW-FPI thus
becomes the most rapidly convergent method among those that leave the
original potential unchanged.  
	
	The error analysis performed in Appendix~E allows us to introduce what we call  ``accelerated'' estimators, which are capable of improving the rate of convergence of any of the aforementioned methods from $\mathcal{O}(1/n^\alpha)$ to $\mathcal{O}(1/n^{\alpha+1})$  
for the first-order correction, and to $\mathcal{O}(1/n^{\alpha+2})$ 
for the second-order correction, respectively.  Although there is a price paid in the 
form of an increase in the variance of the respective estimators, the 
first-order correction appears suitable for general applications. 

\section{Series representations of the Brownian bridge}

The most general series representation of the Brownian bridge is given 
by the Ito-Nisio theorem, the explicit statement of which is presented in Appendix~A.  
We begin by assuming that we are given $\{\lambda_k(\tau)\}_{k \geq 1}$,
a system of functions on the interval $[0,1]$, which, together with the constant function,
 $\lambda_0(\tau)=1$, make up an orthonormal basis in $L^2[0,1]$. If $\Omega$ 
is the space of infinite sequences $\bar{a}\equiv(a_1,a_2,\ldots)$ and 
\begin{equation}
\label{eq:2}
P[\bar{a}]=\prod_{k=1}^{\infty}\mu(a_k)
\end{equation}
 is the (unique) probability measure on $\Omega$ such that the coordinate maps $\bar{a}\rightarrow a_k$ are independent identically distributed variables with distribution probability
\begin{equation}
\label{eq:3}
\mu(a_k\in A)= \frac{1}{\sqrt{2\pi}}\int_A e^{-z^2/2}\,\ud z
\end{equation}
then,
\begin{equation}
\label{eq:4}
B_u^0(\bar{a})\stackrel{d}{=} \sum_{k=1}^{\infty}a_k\Lambda_{k}(u),\; 0\leq u\leq1
\end{equation}
i.e., the right-hand side random series is equal in distribution to a standard Brownian bridge. 
Therefore, the notation~$B_u^0(\bar{a})$ in~(\ref{eq:4}) is appropriate
and allows us to interpret the Brownian bridge as a collection of random
functions of argument~$\bar{a}$, indexed by~$u$.

 Using the Ito-Nisio representation of the Brownian bridge, 
the Feynman-Ka\c{c} formula~(\ref{eq:1}) takes the form
\begin{eqnarray}
\label{eq:5}
\frac{\rho(x,x';\beta)}{\rho_{fp}(x,x';\beta)}&=&\int_{\Omega}\ud P[\bar{a}]\nonumber  \exp\bigg\{-\beta \int_{0}^{1}\! \!  V\Big[x_{0}(u)+ \\
&&\sqrt{\frac{\beta\hbar^2}{m_0}} \sum_{k=1}^{\infty}a_k \Lambda_k(u)
\Big]\ud u\bigg\}.\quad
\end{eqnarray}
To reinforce the formula~(\ref{eq:5}), consider the 
functions~$\{\sqrt{2}\cos(k\pi \tau)\}_{k\geq 1}$, which, together with the constant 
function, make up a complete orthonormal system of~$L^2[0,1]$. Since
\[
\int_{0}^{u}\sqrt{2}\cos(k\pi \tau)\ud \tau = \sqrt{\frac{2}{\pi^2}}\frac{\sin(k\pi u)}{k},
\]
the Ito-Nisio theorem implies that 
\begin{equation}
\label{eq:6}
B_u^0 \stackrel{d}{=} \sqrt{\frac{2}{\pi^2}}\sum_{k=1}^{\infty}a_k\frac{\sin(k\pi u)}{k},\; 0\leq u\leq1,
\end{equation}
so that the Feynman-Ka\c{c} formula~(\ref{eq:5}) becomes
\begin{eqnarray}
\label{eq:7}
\frac{\rho(x,x';\beta)}{\rho_{fp}(x,x';\beta)}&=&\int_{\Omega}\ud P[\bar{a}]\nonumber  \exp\bigg\{-\beta  \int_{0}^{1}\! \!  V\Big[x_{0}(u)+\\ && \sum_{k=1}^{\infty}a_k \sigma_k \sin(k\pi u) \Big]\ud u\bigg\},\quad
\end{eqnarray}
where \[\sigma_k^2=\frac{2\beta\hbar^2}{m_0\pi^2}\frac{1}{k^2}.\]
Equation (7), derived here as a special case of the
Ito-Nisio theorem, is the so-called Fourier path integral method.\cite{Dol84} 
Historically, the sine-Fourier  representation was one of the first explicit constructions 
of the Brownian motion.\cite{Wie23} Following the mathematical literature, 
we shall call it the Wiener construction after the name of its author, even though the 
original FPI method was deduced using arguments other than those presented here.

 The ``primitive'' series representation method consists of approximating the Brownian 
bridge by the $n$-th order partial sum of the series~(\ref{eq:4}). Thus, 
\begin{eqnarray}
\label{eq:8}
\frac{\rho_{P}^n(x,x';\beta)}{\rho_{fp}(x,x';\beta)}&=&\int_{\Omega}\ud P[\bar{a}]\nonumber  \exp\bigg\{-\beta \int_{0}^{1}\! \!  V\Big[x_{0}(u)+ \\
&&\sqrt{\frac{\beta\hbar^2}{m_0}} \sum_{k=1}^{n}a_k \Lambda_k(u) \Big]\ud u\bigg\}\quad
\end{eqnarray} 
An immediate question arises: What is the best choice 
of functions $\lambda_i(u),\; i \geq 1$, independent of potential, such 
that~(\ref{eq:8}) has the fastest convergence? Although the phrase ``independent 
of potential'' carries ambiguities, in the remainder of 
this section we shall provide a more precise statement of the problem.

	We start with the observation that the Wiener basis 
is the only basis for which {\em both}  $\lambda_i(u)$, and their
primitives 
$\Lambda_i(u), \; i \geq 1$ are orthogonal. Indeed, let us notice that by 
construction, $\Lambda_i(0)=0$ for $i \geq 1$ and that 
\[
\Lambda_i(1)= \int_0^1 \lambda_i(\tau)\lambda_0(\tau)\ud \tau =0 \quad \forall \; i \geq 1
\]
by orthogonality and the fact that $\lambda_0(\tau)=1$. The unique basis,
$\Lambda_i(u)$, for which
\begin{eqnarray*}
\int_0^1 \Lambda_i(\tau)\Lambda_j(\tau)\ud \tau =0, && \forall\; i\neq j \\
\int_0^1 \lambda_i(\tau)\lambda_j(\tau)\ud \tau =\delta_{ij},&&  \forall \; i,j \geq 1\\
\Lambda_i(0)=\Lambda_i(1)=0, &&\forall \; i \geq 1
\end{eqnarray*}
is made (up to a multiplication factor) of the eigenfunctions of the Dirichlet problem:
\begin{eqnarray*}
-\frac{1}{2}\Delta \Lambda_i(u)= e_i \Lambda_i(u),\quad \Lambda_i(0)=\Lambda_i(1)=0,
\end{eqnarray*}
as follows from the associated Dirichlet variational principle and the non-degeneracy of the spectrum of the ``particle in a box problem''. But that basis is precisely the Wiener basis.

The orthogonality of the primitives, $\lambda_i(u)$, suggests that the 
Wiener basis is (in a sense that will be made clear below) optimal for the 
representation of the Brownian bridge. Let us define 
\[S_{u}^n(\bar{a})= \sum_{k=1}^{n}a_k \Lambda_k(u)\quad \text{and}\quad B_{u}^n(\bar{a})=
\sum_{k=n+1}^{\infty}a_k \Lambda_k(u),\] as the $n$-th order partial sum in~(\ref{eq:4})
and the corresponding ``tail" series, respectively.  In terms of these
sums, the Brownian bridge is expressed as
 $B_u^0(\bar{a})=S_u^n(\bar{a})+B_u^n(\bar{a})$. 
Obviously, $B_u^n$ and $S_u^n$ are independent. Moreover, 
a standard theorem regarding the sum of independent Gaussian 
distributed random variables shows that $B_u^n$ and $S_u^n$ are 
again Gaussian distributed random variables of mean zero and variances
\[\mathbb{E} (B_u^n)^2=\sum_{k=n+1}^\infty \Lambda_k(u)^2 \quad \text{and} 
\quad \mathbb{E} (S_u^n)^2=\sum_{k=1}^n \Lambda_k(u)^2, \] respectively. 
By independence, we have the equality
\begin{equation}
\label{eq:9}
\mathbb{E} (B_u^0)^2 =\mathbb{E} (B_u^n)^2+ \mathbb{E} (S_u^n)^2 =u(1-u),
\end{equation} where we used the fact that the variance of the Brownian
bridge does  not depend upon the series representation and so, it can be
computed by using any  convenient basis (e.g. the sine-Fourier basis). 

A natural way of measuring the quality of the approximation 
$S_u^n(\bar{a})\approx B_u^0(\bar{a})$ is the value of the time average of 
the variances of the tails 
\begin{eqnarray}
\label{eq:10}
\int_0^1 \mathbb{E} (B_u^0 -S_u^n)^2 \ud u = \int_0^1 \mathbb{E} (B_u^n )^2 \ud u = 
\nonumber \\ \int_0^1 \Big[u(1-u)- \sum_{k=1}^n \Lambda_k(u)^2\Big]\ud u.
\end{eqnarray}
Intuitively, the best approximating series is the one that minimizes the 
functional~(\ref{eq:10}) for each~$n$ (we shall show that the answer is indeed a series). 
More clearly, we want to find $\{\lambda_k(\tau);k \in \overline{1,n}\}$, the 
system of functions on the interval $[0,1]$ which, together with the constant 
function $\lambda_0(\tau)=1$, make up an orthonormal system in $L^2[0,1]$ and 
which realizes the maximum of the functional
\begin{equation}
\label{eq:11}
G(\lambda_1,\ldots,\lambda_n)=\sum_{k=1}^{n}\int_{0}^{1}\Lambda_k^2(u)\ud u.
\end{equation}
Since the system $\{\sqrt{2}\cos(k\pi \tau)\}_{k\geq 1}$ together with the constant 
function make up a complete orthonormal system of~$L^2[0,1]$, we may 
write
\[
\lambda_k(u)=\sum_{l=1}^\infty \sqrt{2}\cos(l\pi u)\int_0^1 \lambda_k(\tau)\sqrt{2}\cos(l\pi \tau) \ud \tau.
\]
Replacing this in~(\ref{eq:11}), we obtain
\begin{widetext}
\begin{eqnarray}
\label{eq:12}
G(\lambda_1,\ldots,\lambda_n)=\sum_{k=1}^{n}\int_{0}^{1}\Lambda_k^2(u)\ud u= \sum_{k=1}^{n} \sum_{l=1}^{\infty}\sum_{j=1}^\infty 2 \int_{0}^{1} \int_{0}^{1} \lambda_k(\tau)\lambda_k(\theta)\cos(l\pi \tau) \cos(j\pi \theta) \ud \tau  \ud \theta \times \nonumber \\
\int_0^1 \frac{2}{\pi^2}\frac{\sin(l\pi u)}{l}\frac{\sin(j \pi u)}{j} \ud u=  \sum_{k=1}^{n}  \int_{0}^{1} \int_{0}^{1} \lambda_k(\tau)\lambda_k(\theta)\sum_{l=1}^{\infty} \frac{2}{\pi^2}\frac{\cos(l\pi \tau) \cos(l\pi \theta)}{l^2} \ud \tau  \ud \theta, 
\end{eqnarray}
\end{widetext}
where we used the fact that the system $\{\sqrt{2}\sin(k\pi \tau)\}_{k\geq 1}$ is also orthonormal. From the theory of integral equations with symmetric kernels, we learn that the maximum of~(\ref{eq:12}) is realized on the set of the~$n$ eigenfunctions having the largest eigenvalues. Since the kernel is already in the series representation form, the maximum of our problem is $\sum_{k=1}^n1/(\pi k)^2$ and is attained on the (orthonormal) functions
\[ 
\lambda_k(u)=\sqrt{2}\cos(k\pi u)\quad k \in \overline{1,n}.
\]
It follows that the Wiener  representation is the unique series for which the time-average of the variance of the tail series reaches the minimum value of
\begin{equation}
\label{eq:13}
\int_0^1 \mathbb{E} (B_u^0 -S_u^n)^2 \ud u = \frac{1}{6}- \sum_{k=1}^n\frac{1}{\pi^2k^2}.
\end{equation} 

However, there is a direct connection between the asymptotic 
rate of convergence of the primitive method and the
quantity~$\mathbb{E}(B_u^0-S_u^n)^2$, a connection that is given by
formula~(\ref{eq:20}) and is analyzed in Section~IIIA. It allows us to
conclude that the Wiener representation is the best series for general
use in the primitive method.

\section{Improvements in the Primitive Fourier Path integral Technique}

   In the primitive series approach [c.f. Eq. (8)], the ``tail'' portion of the
Brownian bridge is simply discarded.  Rather than neglecting
these terms entirely, it is possible to include (approximately) their
effects through a number of approaches.  One of these is known as the
partial averaging  method.\cite{Dol85}  Another is a method we term the 
reweighted method introduced in Section IIIB. We note that in both
methods the
$n$-th order partial sum~$S^n_u$ is unchanged, its distribution being
identical to the primitive method one. All methods which preserve the
distribution of the partial sum~$S^n_u$ are referred to by the name of
the respective series. As such, if the sine-Fourier basis is utilized, we
shall call the aforementioned approaches the PA-FPI and the RW-FPI
methods, respectively. 

\subsection{Partial Averaging Method}
Developed initially for the Fourier path integral method,
the partial averaging technique can be defined for all series
representations.
 The key is the independence of the coordinates $a_k$, which physically amounts to choosing those representations for which the kinetic energy operator is diagonal. 
Denoting by $\mathbb{E}_{n}$ the average over the coefficients beyond the 
rank~$n$, the partial averaging formula reads:
\begin{widetext}
\begin{equation}
\label{eq:14}
\frac{\rho^n_{PA}(x,x';\beta)}{\rho_{fp}(x,x';\beta)}=\int_{\mathbb{R}}\ud \mu(a_1)\ldots \int_{\mathbb{R}}\ud \mu(a_n) \exp\bigg\{-\beta \; \mathbb{E}_n\int_{0}^{1}\! \! 
V\Big[x_{0}(u)+\sqrt{\frac{\beta\hbar^2}{m_0}} \sum_{k=1}^{\infty}a_k
\Lambda_k(u) \Big]\ud u\bigg\}\quad
\end{equation}
\end{widetext}
As we mentioned before,  the series $ \sum_{k=n+1}^{\infty}a_k \Lambda_k(u)$ 
is again a Gaussian distributed variable of mean zero and variance $\mathbb{E}(B_u^n)^2$.
Using this together with the equality~(\ref{eq:9}), it is not difficult to show that 
formula~(\ref{eq:14}) becomes   
\begin{widetext}
\begin{equation}
\label{eq:16}
\frac{\rho^n_{PA}(x,x';\beta)}{\rho_{fp}(x,x';\beta)}=\int_{\mathbb{R}}\ud \mu(a_1)\ldots \int_{\mathbb{R}}\ud \mu(a_n) \exp\bigg\{-\beta \; \int_{0}^{1}\! \! 
\overline{V}_{u,n}\Big[x_{0}(u)+\sqrt{\frac{\beta\hbar^2}{m_0}}
\sum_{k=1}^{n}a_k \Lambda_k(u) \Big]\ud u\bigg\},\quad
\end{equation}
\end{widetext}
where
\begin{equation}
\label{eq:17}
\overline{V}_{u,n}(x)=\int_{\mathbb{R}}\frac{1}{\sqrt{2\pi\Gamma_{n}^2(u)}} \exp\left[-\frac{z^2}{2\Gamma_{n}^2(u)}\right]V(x+z) \ud z,
\end{equation}
with~$\Gamma_n^2(u)$ defined by
\begin{equation}
\label{eq:15}
\Gamma_{n}^2(u)=\frac{\beta\hbar^2}{m_0}\left[u(1-u)-
\sum_{k=1}^{n}\Lambda_k(u)^2\right].
\end{equation}

There is one property of the partial averaging method of particular note:
an application of Jensen's inequality\cite{Dur96a}  shows that
\begin{widetext}
\begin{eqnarray}
\label{eq:18}
\frac{\rho^{n+1}_{PA}(x,x';\beta)}{\rho_{fp}(x,x';\beta)}=\int_{\mathbb{R}}\ud \mu(a_1)\ldots \int_{\mathbb{R}}\ud \mu(a_n)\int_{\mathbb{R}}\ud \mu(a_{n+1}) \exp\bigg\{-\beta \int_{0}^{1}\! \! \nonumber
\overline{V}_{u,n+1}\Big[x_{0}(u)+\sqrt{\frac{\beta\hbar^2}{m_0}}
\sum_{k=1}^{n+1}a_k \Lambda_k(u) \Big]\ud u\bigg\}\geq \\ 
\int_{\mathbb{R}}\ud \mu(a_1)\ldots \int_{\mathbb{R}}\ud \mu(a_n) \exp\bigg\{-\beta \int_{\mathbb{R}}\ud \mu(a_{n+1}) \int_{0}^{1}\! \! 
\overline{V}_{u,n+1}\Big[x_{0}(u)+\sqrt{\frac{\beta\hbar^2}{m_0}}
\sum_{k=1}^{n+1}a_k \Lambda_k(u) \Big]\ud u\bigg\}=\\
\int_{\mathbb{R}}\ud \mu(a_1)\ldots \int_{\mathbb{R}}\ud \mu(a_n) \exp\bigg\{-\beta \int_{0}^{1}\! \! \nonumber
\overline{V}_{u,n}\Big[x_{0}(u)+\sqrt{\frac{\beta\hbar^2}{m_0}}
\sum_{k=1}^{n}a_k \Lambda_k(u) \Big]\ud
u\bigg\}=\frac{\rho^n_{PA}(x,x';\beta)}{\rho_{fp}(x,x';\beta)}.
\end{eqnarray}
\end{widetext}
Therefore, the sequence
\begin{equation}
\label{eq:19}
\rho_{PA}^0(x,x';\beta) \leq \rho_{PA}^1(x,x';\beta) \leq \ldots \leq \rho_{PA}^n(x,x';\beta) \leq \ldots
\end{equation}
is an increasing sequence that converges from below to the true 
density matrix, $\rho(x,x';\beta)$.

Let us now consider the problem of choosing the best series representation 
for use within the partial averaging framework. We notice that the 
sequence $\Gamma_n^2(u)$, given by formula~(\ref{eq:15}), decreases 
monotonically while $\rho_{PA}^n(x,x';\beta)$ increases monotonically, 
as shown by formula~(\ref{eq:19}). In fact, there is a connection 
between~(\ref{eq:15}) and~(\ref{eq:19}) in the sense that the 
faster the Gaussian spread converges to zero, the faster~$\overline{V}_{u,n}(x)$ 
converges to the original potential~$V(x)$, and the faster $\rho_{PA}^n(x,x';\beta)$ 
increases to $\rho(x,x';\beta)$. We note that this observation is general, 
independent of the potential $V(x)$.  Of course, one may try to optimize 
$\rho_{PA}^n(x,x';\beta)$ directly, but then the best basis will depend upon 
the potential, an undesirable computational feature.  We thus
conclude that the optimal basis for the partial averaging method
 is the one for which the time-average 
of $\Gamma_n^2(u)$ has the fastest decrease to zero, $i.e.$ the Wiener or Fourier basis.  
In this sense, the best partial averaging method is the PA-FPI approach.

		We now present one final argument in favor of the Wiener basis, an argument that will 
lead us to a new computational approach, the reweighted FPI technique. 
Remembering the primitive random series method~(\ref{eq:8}) and defining
\begin{widetext}
\[
X_\infty(x,x',\bar{a};\beta)=\exp\left\{-\beta\int_{0}^{1}\! \! 
V\Big[x_{0}(u)+\sqrt{\frac{\beta\hbar^2}{m_0}} \sum_{k=1}^{\infty}a_k
\Lambda_k(u) \Big]\ud u\right\},
\]
and
\[
X_n(x,x',\bar{a};\beta)=\exp\left\{-\beta\int_{0}^{1}\! \! 
V\Big[x_{0}(u)+\sqrt{\frac{\beta\hbar^2}{m_0}} \sum_{k=1}^{n}a_k
\Lambda_k(u) \Big]\ud u\right\},
\]
respectively, we have to the first-order in~$\beta$:
\begin{eqnarray}
\label{eq:20}
&&\mathbb{E}_n X_\infty(x,x',\bar{a};\beta)-X_n(x,x',\bar{a};\beta) \approx \beta
X_n(x,x',\bar{a};\beta)\times \nonumber\\&&
\int_{0}^{1}\!\left\{ \!  V\Big[x_{0}(u)+\sqrt{\frac{\beta\hbar^2}{m_0}}
\sum_{k=1}^{n}a_k \Lambda_k(u) \Big]-
\overline{V}_{u,n}\Big[x_{0}(u)+\sqrt{\frac{\beta\hbar^2}{m_0}}
\sum_{k=1}^{n}a_k \Lambda_k(u) \Big]\right\}\ud u.
\end{eqnarray}
\end{widetext}
The rate of convergence for the primitive random series method thus
depends on the  difference between $V(x)$ and $\overline{V}_{u,n}(x)$,
which in turn depends on the value  of $\Gamma_n^2(u)$. Therefore, to a
first approximation, the differences between the exact  and the~$n$-th
order FPI density matrices depend not on the detailed structure of the
respective tails, but rather on the spread of the tail series, 
$B_u^n(\bar{a})$, a quantity whose time average reaches a minimum for the Wiener series.  One can readily
verify that the term of order~$\beta$ vanishes for the partial averaging
analog of formula~(\ref{eq:20}), an indication that the technique exactly 
accounts for the extra spread of the paths induced by the tail series.

\subsection{Reweighting Method}

Unlike the partial averaging method, the reweighting technique attempts to
account for the effects of the tail series in a way that does not
involve modifying the associated potential energy.  We shall work out the
result for the Wiener basis, noting that: (1) the approach can  be
applied to any arbitrary representation, and (2) the efficiency  of the
method will depend upon the  specific series selected.  The basic idea is
to replace
$B_u^n(\bar{a})$  by another collection, $R_u^n(b_1,\cdots,b_n)$, which
is supported by an $n$-dimensional  underlying probability space.  We
require that:
\begin{enumerate}
\item{The variance at the point~$u$ of $R_u^n(b_1,\cdots,b_n)$, 
denoted by ${\Gamma'}^2_n(u)$, be as close as possible to $\Gamma_n^2(u)$. }
\item{The variables $S_u^n(a_1,\cdots, a_n)$ and $R^n_u(b_1,\cdots,b_n)$ be 
independent and their sum have a joint distribution as close to a Brownian 
bridge as possible. }  
\end{enumerate}

 One possible candidate for our approach is to choose 
$R_u^n(b_1,\cdots,b_n)=\sqrt{\frac{\beta\hbar^2}{m_0}}\sum_{k=1}^n b_k
\Omega_k(u),$  with $b_1,\cdots,b_n$ independent identically distributed
standard normal random  variables. Condition~2 above is realized in the 
Ito-Nisio theorem  by insuring that the collection $\{\cos(k\pi u), \omega_k(u)\}_{ k \geq 1}$ 
is orthogonal, where $\omega_k(u)$ is the derivative of $\Omega_k(u)$.  We shall enforce this condition by choosing 
$\Omega_k(u)=\alpha_{n,k}\sin[(k+n)\pi u]$ where $\alpha_{n,k}$ are some constants 
yet to be determined. With the condition~1 above in mind, and by noticing that 
in the exact FPI representation~(\ref{eq:7}) 
the terms of the form $\sin[(k+n j)\pi u]$ with $j \geq 1$ ``decouple'' as $n \rightarrow \infty$, our intuition tells us that a good candidate for $\alpha_{n,k}$ is 
\begin{equation*}
\alpha_{n,k}^2=\frac{2}{\pi^2}\sum_{j=1}^\infty \frac{1}{(k+j n)^2}.
\end{equation*}  
With this choice, the $n$-th order RW-FPI density matrix is given by the formula
\begin{eqnarray}
\label{eq:21}
\frac{\rho_{RW}^n(x,x';\beta)}{\rho_{fp}(x,x';\beta)}&=&\int_{\Omega}\ud P[\bar{a}]\nonumber  \exp\bigg\{-\beta \int_{0}^{1}\! \!  V\Big[x_{0}(u)+ \\ && + \sum_{k=1}^{2n}a_k \sigma_{n,k} \sin(k\pi u) \Big]\ud u\bigg\},\quad
\end{eqnarray} 
where 
\begin{equation}
\label{eq:22}
\sigma_{n,k}^2=\frac{2\beta\hbar^2}{m_0\pi^2}\times \left\{\begin{array}{ll}1/k^2,& \text{if}\; 1\leq k\leq n\\ \\  \sum_{j=0}^\infty 1/(k+jn)^2, &  \text{if}\; n < k \leq 2n.
\end{array}\right.
\end{equation}
The evaluation of the path weights $\sigma_{n,k}^2$ is discussed in 
Appendix~D.

Clearly, our choice of $R^n_u(b_1,\cdots,b_n)$ is not unique. For a better understanding 
of the quality of the approximation, let us compare numerically:
\begin{itemize}
\item{$\Gamma_n^2(u)$, the tail variance for the full FPI representation and for the PA-FPI method,} 
\item{${\Gamma'}^2_n(u)=\sum_{k=n+1}^{2n}\sigma_{n,k}^2\sin(k\pi u)^2$, the tail variance for the RW-FPI method, and}
\item{${\Gamma''}^2_n(u)=\sum_{k=n+1}^{2n}\sigma_{k}^2\sin(k\pi u)^2$, the 
tail variance for the  FPI method if it were computed without reweighting
($i.e.$ by simply  considering the next~$n$ Fourier terms).}
\end{itemize}
Fig~1 plots the above variances for $n=9$. We notice that 
$\Gamma_n^2(u)$ and ${\Gamma'}^2_n(u)$ are indeed close, much closer than 
the result obtained by simply expanding the primitive FPI approach with a 
similar number
of additional terms.  

\begin{figure}[!tbp] 
  \includegraphics[clip=t]{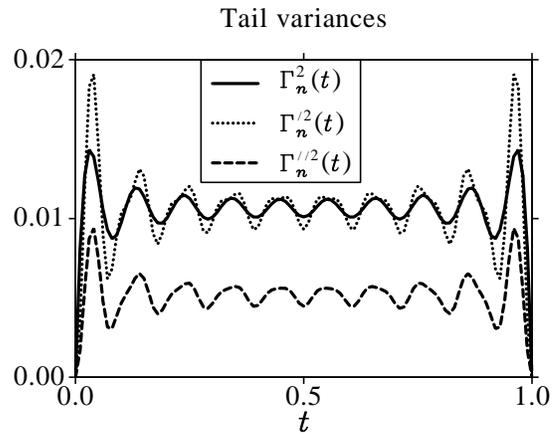}
 \caption[sqr]
{\label{Fig:1}
A plot of the tail variances for the PA-FPI, RW-FPI, and $2n$-order
primitive FPI for $n=9$. Notice that the simple inclusion of the next~$9$
terms within the primitive FPI is not the optimal strategy. }
\end{figure}

\section{Asymptotic convergence of the FPI techniques}

	We say that a given method converges asymptotically as $\mathcal{O}(1/n^\alpha)$ 
if the partition function, the density matrix at each pair of points $(x,x')$, and 
their first-order temperature derivatives converge as fast as $\mathcal{O}(1/n^\alpha)$.  
Generally speaking, the aforementioned quantities may have different asymptotic rates of 
convergence.  However, if the potential is smooth enough, our intuition says that 
this is not true.  For the case of the harmonic oscillator, we shall only 
verify the convergence of the partition function. On the other hand, 
for numerical simulations it is more convenient to compute the average energy of 
the system with the help of the so called T-estimator, which can solely be expressed 
as a functional of the diagonal density matrix:
\begin{equation}
\label{eq:23}
\left\langle E \right\rangle^{T}_{\beta} =-\frac{\partial}{\partial \beta} \ln{\left[ \int_{\mathbb{R}} \rho(x;\beta) \ud x \right]}. 
\end{equation}
The above formula can be expressed as the statistical average
\begin{equation}
\label{eq:24}
\left\langle E \right\rangle^{T}_{\beta} =\frac{\int_{\mathbb{R}}\ud x\int_{\Omega}\ud P[\bar{a}] X_n(x,\bar{a};\beta)E_n^T(x,\bar{a};\beta) }{\int_{\mathbb{R}}\ud x\int_{\Omega}\ud P[\bar{a}] X_n(x, \bar{a};\beta)},
\end{equation}
which can be evaluated by Monte Carlo integration. 
Using the notation
\[ \ x_n(\bar{a},u;\beta)=\sum_{k=1}^{n}a_k\sigma_k \sin(k \pi u)\] to denote the stochastic portions
of the current path truncated to the first~$n$ terms, one easily deduces that the T-estimator function for the
primitive FPI method is
\begin{widetext}
\begin{eqnarray}
\label{eq:25}
 E_n^T(x, \bar{a};\beta)&=&\frac{1}{2\beta}+ \int_{0}^{1}\! \! V[x+ x_n(\bar{a},u;\beta) ]\,\ud u +
\frac{1}{2}\int_{0}^{1}\! \!  V'[x+ x_n(\bar{a},u;\beta)]\ x_n(\bar{a},u;\beta)\,\ud u, 
\end{eqnarray}
while for the PA-FPI method, one obtains
\begin{eqnarray}
\label{eq:26}
 E_n^T(x, \bar{a};\beta)&=&\frac{1}{2\beta}+ \int_{0}^{1}\! \!  \overline{V}_{u,n}[x+ x_n(\bar{a},u;\beta) ]\,\ud u +
\frac{1}{2}\int_{0}^{1}\! \!  \overline{V}'_{u,n}[x+ x_n(\bar{a},u;\beta)]\ x_n(\bar{a},u;\beta)\,\ud u + \nonumber \\ &&
\frac{1}{2}\int_{0}^{1}\! \!  \overline{V}''_{u,n}[x+ x_n(\bar{a},u;\beta)]\,\Gamma_n^2(u)\,\ud u.
\end{eqnarray}
By a simple integration by parts against the coordinate~$x$, one may eliminate 
the second derivative of the potential and obtain the following equivalent PA-FPI energy 
estimator:
\begin{eqnarray}
\label{eq:27}
 E_n^T(x, \bar{a};\beta) &=&\frac{1}{2\beta}+ \int_{0}^{1}\! \!  \overline{V}_{u,n}[x+ x_n(\bar{a},u;\beta) ]\,\ud u +
\frac{1}{2}\int_{0}^{1}\! \!  \overline{V}'_{u,n}[x+ x_n(\bar{a},u;\beta)]\ x_n(\bar{a},u;\beta)\,\ud u + \nonumber \\&&
\frac{\beta}{2}\left\{\int_{0}^{1}\! \!  \overline{V}'_{u,n}[x+ x_n(\bar{a},u;\beta)]\,\Gamma_n^2(u)\,\ud u\right\}\left\{ \int_{0}^{1}\! \!  \overline{V}'_{u,n}[x+x_n(\bar{a},u;\beta)]\,\ud u\right\}.
\end{eqnarray}

	The T-estimator for the RW-FPI method has the same expression as the one for 
primitive FPI, except for the redefinition of the current 
path \[ \ x_n(\bar{a},u;\beta)= \sum_{k=1}^{2n} a_k \sigma_{n,k}\sin(k \pi u).\]	

It is important to note that because of the way we have included
 the temperature dependence of the path distribution in the above analysis, 
we have obtained directly the so called ``virial'' forms of the energy estimators.
These virial expressions have desirable variance properties\cite{Ele98, Fre84} and 
are generally preferred for precise Monte Carlo applications.  
The special form~(\ref{eq:27}) of the PA-FPI energy estimator is
numerically advantageous  since it does not require the evaluation of the
second derivatives of the averaged  potential. Although we do not study
it in this paper because it is not  a functional of the diagonal density
matrix, the H-estimator for the PA-FPI method can similarly be put in the
simple form: 
\begin{equation}
\label{eq:28}
 E_n^H(x, \bar{a};\beta)=\frac{1}{2\beta}+ V(x)+\frac{\hbar^2 \beta^2}{4 m_0}
\int_{0}^{1}\! \! \int_{0}^{1}\! \!(u-\tau)^2 \overline{V}'_{u,n}[x+x_n(\bar{a},u;\beta)]\,\overline{V}'_{\tau,n}[x+x_n(\bar{a},\tau;\beta)]\,\ud u \,\ud \tau.
\end{equation}
The equivalent H-estimator functions for the primitive FPI and RW-FPI
approaches
 look formally the same, 
except that the potential is no longer averaged. The H-estimator is thus
properly defined even for potentials that do not have second-order derivatives. 
The reader should notice that the double integral appearing in~(\ref{eq:28}) is really 
a sum of products of monodimensional integrals. We chose this representation for symmetry 
purposes. The estimator is thus the sum of the ``classical'' energy and 
a ``quantum'' correction term.
\end{widetext}
\subsection{Partition functions for the harmonic oscillator} 
In Ref.~\onlinecite{Ele98}, enough analytical evidence was presented to suggest that the asymptotic behavior of the primitive and partial averaging FPI methods is controlled
at most by the values of the second derivatives of the potential. Here, we conjecture that this remains true of the RW-FPI method, so that an analysis of the harmonic oscillator, the simplest potential having a non-vanishing second-order derivative, should provide a reliable guess of the asymptotic rates for all ``smooth'' potentials (defined here as the potentials having continuous second-order derivatives). Therefore, we shall study the asymptotic convergence of the partition function for a one-dimensional particle of mass $m_0=1$ moving in the quadratic potential $V(x)=x^2/2$. We also set $\hbar=1$ and $\beta=1$.

The exact analytical expressions for the harmonic oscillator 
partition functions are derived in the Appendix~B for the 
three methods: primitive FPI, PA-FPI, and RW-FPI, respectively. 
The partition functions of even and odd orders have a slightly different 
convergence behavior according to whether $1-(-1)^n$ is~$0$ or~$2$ 
(see Appendix~B). To avoid the appearance of certain oscillations 
in our plots, we shall only compute the odd subsequence for the RW-FPI method. 
Remember, however, that the $2n+1$-th order RW-FPI approach uses in fact
twice as many points.  To ensure fairness as far as the computational
effort is concerned,  we shall compare the $2n+1$-th order RW-FPI results
with those of the $4n+2$-th order  primitive FPI and PA-FPI approaches
since, for a given order, the former
 method uses twice as many path variables as do the latter techniques. 
It is convenient to redefine the order of the RW-FPI 
method as being equal to the number of random variables used to parameterize the paths, 
in this case:~$4n+2$.  In general, 
\begin{eqnarray}
\label{eq:29}
\frac{\rho_{RW}^{2n}(x,x';\beta)}{\rho_{fp}(x,x';\beta)}&=&\int_{\Omega}\ud P[\bar{a}]\nonumber  \exp\bigg\{-\beta \int_{0}^{1}\! \!  V\Big[x_{0}(u)+ \\ && + \sum_{k=1}^{2n}a_k \sigma_{n,k} \sin(k\pi u) \Big]\ud u\bigg\},\quad
\end{eqnarray} 
where $\sigma_{n,k}$ is given by formula~(\ref{eq:22}) and is evaluated in Appendix~D. 

	Let us assume that we may expand the difference $Z^{n}_{Pr}(\beta)-Z(\beta)$ as the generalized power series
\[Z^{n}_{Pr}(\beta)-Z(\beta)=\frac{c}{n^\alpha}\left(1+\sum_{k=1}^\infty\frac{c_k}{n^k}\right),\]
with $c \neq 0$. For~$n$ large enough, it suffices to consider the approximation
\begin{equation}
\label{eq:30}
Z^{n}_{Pr}(\beta)-Z(\beta)\approx\frac{c}{n^\alpha}\left(1+\frac{c_1}{n}\right).
\end{equation}
By passing to the subsequence~$4n+2$ and taking the ratios of consecutive 
differences, we obtain:
\[\frac{Z^{4n-2}_{Pr}(\beta)-Z(\beta)}{Z^{4n+2}_{Pr}(\beta)-Z(\beta)}\approx\left(\frac{4n+2}{4n-2}\right)^\alpha\frac{1+c_1/(4n+2)}{1+c_1/(4n-2)}.\]
Next, we take the logarithm and use the approximation $1/(1+x)\approx 1-x$ for the last term: 
\begin{eqnarray*}
\log\left(1+\frac{Z^{4n-2}_{Pr}(\beta)-Z^{4n+2}_{Pr}(\beta)}{Z^{4n+2}_{Pr}(\beta)-Z(\beta)}\right)\approx \\ \alpha \log\left(1+\frac{4}{4n-2}\right)+\log\left(1-\frac{c_1}{4n^2-1}\right).
\end{eqnarray*}
We expand the logarithms on the right-hand side of the above equation so that the error be of the order $\mathcal{O}(1/n^3)$ and then multiply the resulting equation by $n^2-1/4$ to obtain 
\begin{eqnarray*}
(n^2-1/4)\log\left(1+\frac{Z^{4n-2}_{Pr}(\beta)-Z^{4n+2}_{Pr}(\beta)}{Z^{4n+2}_{Pr}(\beta)-Z(\beta)}\right)\approx \\ n \alpha +\alpha/2-\alpha\frac{4n+2}{4n-2}-c_1/4.
\end{eqnarray*}
It is convenient to introduce the notation
\[
DZ^{4n+2}_{Pr}(\beta)=Z^{4n-2}_{Pr}(\beta)-Z^{4n+2}_{Pr}(\beta)
\]
and set
\[
\alpha_{Pr}^n=(n^2-1/4)\log\left(1+\frac{DZ^{4n+2}_{Pr}(\beta)}{Z^{4n+2}_{Pr}(\beta)-Z(\beta)}\right).
\]
Since $(4n+2)/(4n-2)\approx 1$ for~$n$ large, we conclude
\begin{equation}
\label{eq:31}
\alpha_{Pr}^n\approx  \alpha n-\alpha/2-\frac{c_1}{4},
\end{equation}
which shows that $\alpha_{Pr}^n$ should be asymptotically  a straight line whose slope gives the convergence order. Here, $Z(\beta)$ is the exact value of the partition function and the index~$Pr$ is used to denote the
primitive FPI method. Of course, similar expressions can be written for
the other two methods identified by the indices~$RW$ and~$PA$. For
general expressions which apply to any of the techniques, we shall use
the index~$Mt$.

Once the asymptotic order is established, we may determine the value of the constant~$c$ by analyzing the slope of the equation
\begin{equation}
\label{eq:32}
c_{Mt}^n\approx  cn+c/2+cc_1,
\end{equation}
where 
\[c_{Mt}^n=(4n+2)^{\alpha}{(n+1/2)}\left[Z^{4n+2}_{Mt}(\beta)-Z(\beta)\right].\]
The asymptotic behavior implied by~(\ref{eq:32}) can easily be established by 
replacing~$n$ by~$4n+2$ in equation~(\ref{eq:30}). 

 Fig.~2 shows that the linear region predicted by our analysis is 
quite rapidly reached 
for the harmonic oscillator. One easily notice that the PA-FPI 
and RW-FPI methods 
have similar asymptotic behavior, while the primitive FPI approach has a
slower rate of convergence. 

\begin{figure}[!tbp] 
   \includegraphics[clip=t]{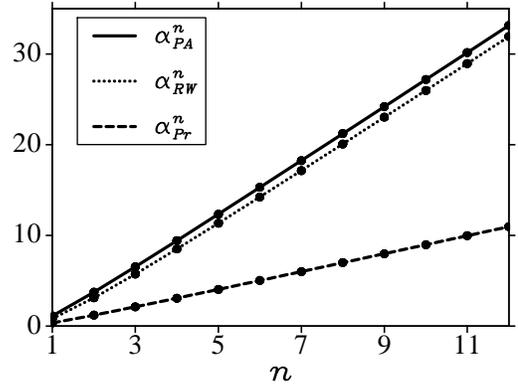}
 \caption[sqr]
{\label{Fig:2}
A plot of the indices of convergence for the PA-FPI, RW-FPI, and
primitive FPI for the quadratic potential.  }
\end{figure}
The asymptotic slopes are computed as the slope of the line that 
best fits the last $[N/3]$ values, where~$N$ is the number of data
points calculated. We assume that we computed enough points so that the
last~$[N/3]$ are in the asymptotic region. Euler least-square fit gives
then the value
\begin{equation}
\label{eq:33}
\alpha_{Mt}=\frac{[N/3]\sum_{k} k\alpha_{Mt}^k-\sum_{k}\alpha_{Mt}^k\cdot\sum_{k} k}{[N/3]\sum_k  k^2-(\sum_{k}k)^2},
\end{equation}
where the summation is done over the last~$[N/3]$ data points. Of course,
the exact value for~$\alpha$ is the limit as $N \rightarrow \infty$ of the
right-hand side of the above formula.    For $N=12$,~(\ref{eq:33}) gives:
$\alpha_{Pr}=1.002$, $\alpha_{PA}=3.007$  and $\alpha_{RW}=3.008$,
suggesting that the asymptotic behavior  is $\mathcal{O}(1/n)$ for the
first, and $\mathcal{O}(1/n^3)$ for the last two methods, respectively. 

The constants~$c$ are calculated in a  similar fashion and the numerical values for 
$N=12$ are:
$c_{Pr}=0.049$, $c_{PA}=-7.933\cdot10^{-3}$, and $c_{RW}=0.887$, respectively. Therefore, the
partial averaging method is superior to the reweighted method in the
sense that it has a smaller convergence constant (smaller in modulus).

Theoretically, if we can compute the difference between successive values of the partition 
function with sufficient precision, we can improve the convergence of any of the FPI 
methods by using better estimators. For first-order, the result can be obtained as 
follows: formula~(\ref{eq:30}) shows that 
\begin{equation}
\label{eq:34}
Z^{4n+2}_{Mt}(\beta)-\frac{c}{(4n+2)^\alpha}
\end{equation}
converges to the exact answer as fast as $\mathcal{O}(1/(4n+2)^{\alpha+1})$ 
and therefore, the last equation is a better estimator as far as the asymptotic 
behavior is concerned. Given that the convergence exponent~$\alpha$ is known, the 
constant~$c$ can be approximately (but arbitrarily exactly as $n \to \infty$) evaluated 
from the equation:
\[
DZ^{4n+2}_{Mt}(\beta)\approx c \frac{(4n+2)^\alpha-(4n-2)^\alpha}{(4n+2)^\alpha (4n-2)^\alpha}\approx  \frac {c}{(4n+2)^\alpha} \frac{\alpha}{n}
\]
Solving for~$c$ and replacing in~(\ref{eq:34}), one ends up with the first-order corrected estimator
\begin{equation}
\label{eq:35}
FZ^{4n+2}_{Mt}(\beta)=Z^{4n+2}_{Mt}(\beta)-\frac{n}{\alpha}DZ^{4n+2}_{Mt}(\beta)
\end{equation}
The second-order estimator can be derived by applying the first-order correction to the first-order estimator. One easily computes:
\begin{eqnarray}
\label{eq:36}
SZ^{4n+2}_{Mt}(\beta)=Z^{4n+2}_{Mt}(\beta)-\frac{(2\alpha+1)n}{\alpha(\alpha+1)}DZ^{4n+2}_{Mt}(\beta)-\nonumber \\
\frac{n}{\alpha(\alpha+1)}DZ^{4n-2}_{Mt}(\beta)+\nonumber \\ \frac{n^2}{\alpha(\alpha+1)}\Big[DZ^{4n-2}_{Mt}(\beta)-DZ^{4n+2}_{Mt}(\beta)\Big]
\end{eqnarray} 
The asymptotic convergence of this estimator is $\mathcal{O}(1/n^{\alpha+2})$.

In principle, one can continue this process beyond second-order.  However, as we shall see in the Appendix~E, such higher order estimators are of little practical value. 
To demonstrate the behavior of the corrected estimators, we compute the convergence exponents for the primitive FPI using the corresponding analog of equation~(\ref{eq:31}). 
Fig.~3 clearly shows the difference in the rate of convergence for the
original  and corrected estimators. The numerical values are
$\alpha_{Z}=1.002$, $\alpha_{FZ}=1.997$,  and $\alpha_{SZ}=2.958$,
demonstrating our predictions. From now on, we shall refer to the 
original, unaccelerated estimator as the zero-order estimator. 
\begin{figure}[!tbp] 
   \includegraphics[clip=t]{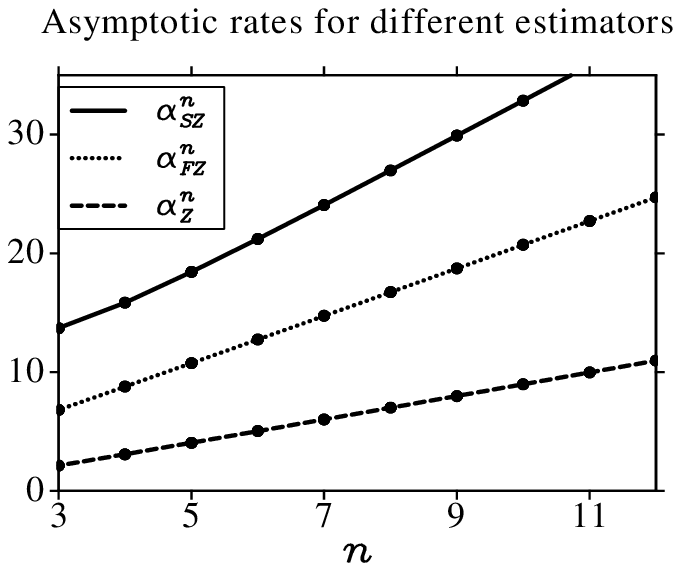} 
 \caption[sqr]
{\label{Fig:3}
A plot of the exponents of convergence for the three $Z$-estimators. The method employed is
primitive FPI as applied to the quadratic potential.  }
\end{figure}

\subsection{A numerical example: The quartic potential}

As we said in the beginning of this section, for numerical purposes 
it is convenient 
to study the convergence of the T-method energy estimator in the 
virial form, 
which can be computed by Monte Carlo integration. As we explain 
below, the numerical study of the asymptotic behavior is not a
computationally easy task, especially for those methods that have rapid
asymptotic convergence. More explicitly, let us take a look at the
following analog of~(\ref{eq:31}):
\begin{equation}
\label{eq:37}
\alpha_{Mt}^n\approx  \alpha_{Mt} n-\alpha_{Mt}/2-\frac{c_{1,Mt}}{4},
\end{equation} 
where  
\begin{eqnarray*}
\alpha_{Mt}^n=(n^2-1/4)\log\left(1+\frac{E^{4n-2}_{Mt}-E^{4n+2}_{Mt}}{E^{4n+2}_{Mt}-E}\right).
\end{eqnarray*}
For the partial averaging method, we suggested that the difference $E^{4n+2}_{PA}-E$ 
decays to zero as fast as~$1/n^3$. In turn, the differences $E^{4n-2}_{PA}-E^{4n+2}_{PA}$ 
between consecutive terms decay to zero as fast as~$1/n^4$. It is thus clear 
that faster rates of convergence of the method require greater precision in the 
evaluation of the terms $E^{4n+2}_{PA}$. If we assume an independent sampling of the probability density shown in formula~(\ref{eq:24}), the  error in the Monte Carlo evaluation of $E^{4n+2}_{PA}$ is 
$\Delta E^{4n+2}_{PA}/\sqrt{N}$, where~$N$ is the number of Monte Carlo sampling points and $\Delta E^{4n+2}_{PA}$ is the standard deviation. This error should satisfy the inequality:
\[|\Delta E^{4n+2}_{PA}|/\sqrt{N} \ll |E^{4n-2}_{PA}-E^{4n+2}_{PA}|\]
It follows that the number of Monte Carlo points necessary to insure a given relative 
error for $\alpha_{PA}$ scales at least as badly as $N\propto n^8$ as a
function of  the number of Fourier coefficients. The same is true for
RW-FPI, while for the  primitive FPI we only need $N\propto n^4$. We
emphasize that this scaling is related to our immediate task of
establishing the asymptotic rates of convergence and is not an issue that
would arise in typical numerical applications.

The second observation we make is that the ratio 
\[|E^{4n-2}_{Mt}-E^{4n+2}_{Mt}|/|\Delta E^{4n+2}_{Mt}| \] increases as the temperature 
is dropped.  Consequently, we would like to conduct our model computations at 
low temperature, where the quantum effects are big enough so that the differences between consecutive terms are significant. At high temperature, the classical limit is a good approximation and these differences may be smaller than the statistical errors we are able to achieve. We are therefore forced to conduct our computations in the ``unfavorable'' range of temperatures, and in general, we need to study groundstate problems. 

We hope this is enough rationale to justify the need for 
a special Monte Carlo integration 
scheme capable of accurately sampling the low temperature 
distributions with good 
efficiency and low correlation, at least for certain classes 
of simpler systems. One such scheme is discussed in Appendix~E, and it
generally applies to the class of single-well potentials. 

For comparison purposes, we shall also compute the T-estimator 
energies for the trapezoidal Trotter method. Expressions similar to those
presented here for the FPI methods were deduced by Coalson\cite{Coa86}
and employed by Mielke and Truhlar\cite{Mie01} as the TT-FPI method. We
shall keep this name in the present paper, though, as defined here, the
TT-FPI approach is
\emph{not} an FPI method because the $n$-th order partial sum~$S^n_u$ is
not the one for the primitive FPI. The importance of this method consists of
the fact that its asymptotic rate of convergence is~$\mathcal{O}(1/n^2)$
for smooth enough potentials, being the fastest primitive method to
date that leaves the potential unchanged.\cite{Rae83} We do not present this scheme in the present paper and
for further information we refer the reader to the cited literature.  

The prototype system studied in this work is  the quartic potential $V(x)=x^4/2$. We set $\hbar =1$ and $m_0=1$ and $\beta =10$. The groundstate of the quartic potential was evaluated by variational methods to be $E_0=0.530181$, while the average energy at the temperature corresponding to $\beta=10$ is $E=0.530183$. We computed the average energy for the sequence $4n+2$ with $1 \leq n \leq 12$, 
corresponding to the actual numbers of Fourier coefficients $6, 10, \ldots, 50$. 
In these calculations, the number of points employed in the Gauss-Legendre
quadrature scheme  was $200$. We used $1.25\cdot 10^{8}$ Monte Carlo
points for primitive  FPI, $2.5\cdot 10^8$ for TT-FPI, $5\cdot 10^{8}$
points for RW-FPI,  and $2 \cdot 10^{9}$ points for PA-FPI calculations,
respectively. The values of
$R_{PA}^{4n+2}$ were  previously computed in a quarter of these numbers
during a ``warm-up'' period, but we  continued to improve them during the
main Monte Carlo procedure. Table~\ref{Tab:II} of Appendix~F summarizes
the results of the computer evaluations. The differences between
successive energy terms were computed with the help of the
estimator~(\ref{eq:46}). The errors  were computed with the help of the
formulae~(\ref{eq:40}) for the average energies,  and~(\ref{eq:48}) for
the estimated differences.
\begin{figure}[!tbp] 
   \includegraphics[clip=t]{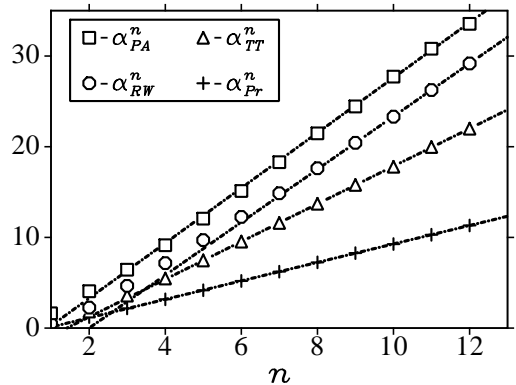} 
 \caption[sqr]
{\label{Fig:4}
The straight lines drawn represent the linear least square fit for the 
last four data. Their slopes give the convergence exponents for each method.
}
\end{figure}  
 
Fig~4 shows the behavior of the functions~$\alpha_{Mt}^n$ for the four methods. 
Among the non-averaged methods, we remark that the primitive FPI approach
reaches its asymptotic  behavior faster than the TT-FPI method, which in
turn reaches its asymptotic region faster than the RW-FPI technique. This
behavior is shown in Fig~5, which plots the current 
slope~$\alpha_{Mt}^{n}-\alpha_{Mt}^{n-1}$. Although the RW-FPI method did
not reach its final asymptotic behavior, the trend is clear. The
computed convergence exponents  using the last four data points are: 
$\alpha_{PA}=3.082$, $\alpha_{RW}=2.917$, $\alpha_{TT}=2.071$, and $\alpha_{Pr}=1.019$.  
Therefore, we conclude that the asymptotic convergence of the methods 
is $\mathcal{O}(1/n^3)$ for PA-FPI and RW-FPI, $\mathcal{O}(1/n^2)$ for TT-FPI, 
and $\mathcal{O}(1/n)$ for primitive FPI. 
\begin{figure}[!tbp] 
   \includegraphics[clip=t]{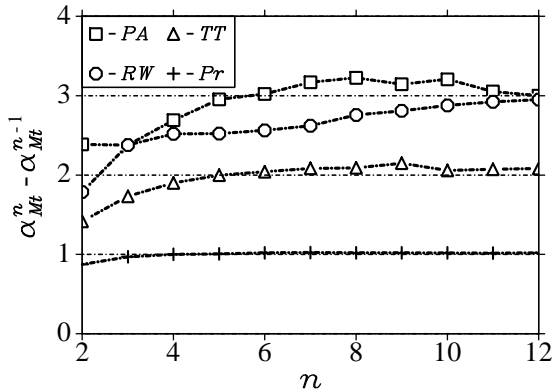} 
 \caption[sqr]
{\label{Fig:5}
The current slopes for each method should ideally converge to~$3$ for PA-FPI 
and RW-FPI, $2$ for TT-FPI, and $1$ for primitive FPI.
}
\end{figure} 
Lastly, it is worth comparing the convergence constants for the PA-FPI 
and RW-FPI methods 
since they have the same asymptotic convergence exponent. The numerical values 
are $c_{PA}=59.4$  and $c_{RW}= -736.7$, showing that the PA-FPI method is 
over~$10$ times faster than the RW-FPI method. This is in 
agreement with the observations made for the partition function of the
harmonic oscillator in the previous section. The PA
speed-up of the convergence is important, especially with respect to
minimizing the number of path variables required in practical
applications.

\section{Conclusions}

In this paper, we have shown that the best series representation 
(with respect to asymptotic convergence) for use in Monte Carlo path
integral methods is the  Wiener sine-Fourier series. Both the RW-FPI and
TT-FPI methods are not series representations and we suggest that the
latter also falls in the category of reweighting techniques.  The partial
averaging technique has the asymptotic convergence $\mathcal{O}(1/n^3)$, 
with a small convergence constant and it is the best way of improving the
asymptotic behavior of the primitive FPI method (at the cost of computing
the Gaussian  transform of the potential). The TT-FPI and RW-FPI methods
increase the order of convergence  of the primitive FPI to
$\mathcal{O}(1/n^2)$, and
$\mathcal{O}(1/n^3)$, respectively, without increasing the variance of
the corresponding estimators. It should be noted that, unlike the
complete partial averaging approach, the reweighting method does not
require the Gaussian transform of the potential.  While a final decision
awaits detailed future studies, we anticipate that this latter feature of
the reweighting approach will be beneficial for applications where the
Gaussian transform is either formally ill-posed and/or computationally
difficult to obtain.  Finally, as discussed in Appendix~E, the first and
the second-order estimators also improve the asymptotic convergence. 
Although both have
larger variances, the first order estimator appears computationally 
feasible since its variance decreases with the number of Fourier
coefficients. 

\begin{acknowledgments}
   The authors acknowledge support from the National Science Foundation through 
awards CDA-9724347, CHE-0095053, and CHE-0131114.  They also would like to
thank Professor D. L. Freeman and Dr. Dubravko Sabo for continuing
discussions concerning the present developments.

\end{acknowledgments}

\appendix
\section{Ito-Nisio theorem}
\begin{1}[Ito-Nisio\cite{Kwa92}]
Let $\{\lambda_k(\tau)\}_{k \geq 0}$ be any orthonormal basis in $L^2[0,1]$ such that $\lambda_0(\tau)=1$, let 
\[
\Lambda_k(u)=\int_{0}^{u}\lambda_k(\tau)\ud \tau, 
\]
and let $\bar{a} \equiv \{a_k\}_{k \geq 1}$ be a sequence of independent identically distributed (i.i.d.) standard normal random variables. Then, the series 
\[
\sum_{k=1}^\infty a_k \Lambda_k(u)
\]
is uniformly convergent almost surely and equal in distribution with a standard Brownian bridge. 
\end{1}

\section{Harmonic oscillator}
\begin{widetext}
The $2n$-th order primitive FPI approximation of the partition 
function for an harmonic oscillator centered at the origin has the
expression 
\[
Z^{2n}_{Pr}(\beta)=\int_{\mathbb{R}} \ud x\int_{\mathbb{R}}\ud a_1 \ldots \int_{\mathbb{R}} \ud a_{2n} \rho^n_{Pr}(x, \bar{a};\beta),
\]
where
\[
\rho^n_{Pr}(x,\bar{a};\beta)=\sqrt{\frac{m_0}{2\pi \hbar^2 \beta}}(2\pi)^{-n}\exp\left(-\sum_{k=1}^{2n} a_k^2/2\right)\exp\left\{-\beta\frac{m_0\omega^2} {2} \int_{0}^{1} \left[ x+ \sum_{k=1}^{2n} a_k \sigma_k \sin(k \pi u)\right]^2\ud u\right\}.
\]
By explicitly computing the integral over~$t$ and then completing the
square, one obtains
\begin{eqnarray}
\label{eq:B1}
\rho^{2n}_{Pr}(x,\bar{a};\beta)=\sqrt{\frac{m_0}{2\pi \hbar^2 \beta}}(2\pi)^{-n}\exp\left\{-\frac{1}{2}\sum_{k=1}^{2n} \left[\xi_k a_k+\frac{\beta m_0 \omega^2 x}{\xi_k} \frac{1-(-1)^k}{k\pi}\right]^2\right\} \times \nonumber
\\ \exp\left\{-\frac{\beta m_0 \omega^2}{2}x^2\left[1-\sum_{k=1}^{2n} \frac{\beta m_0 \omega^2 \sigma_k^2}{\xi_k^2}\left(\frac{1-(-1)^k}{k\pi}\right)^2\right]\right\},
\end{eqnarray}
where
\[\xi_k^2=1+\beta m_0 \omega^2 \sigma^2_k/2.\]
For use as a trial density in Monte Carlo simulations, it is convenient to replace the last factor by its limit $n \to \infty$:
\begin{equation}
\label{eq:B2}
\rho^{2n}_{tr}(x,\bar{a};\beta)=\sqrt{\frac{m_0}{2\pi \hbar^2 \beta}}\exp\left[-\frac{m_0 \omega}{\hbar}x^2\tanh\left(\frac{\hbar\omega}{2}\beta\right)\right]\frac{1}{(2\pi)^{n}}\exp\left\{-\frac{1}{2}\sum_{k=1}^{2n} \left[\xi_k a_k+\frac{\beta m_0 \omega^2 x}{\xi_k} \frac{1-(-1)^k}{k\pi}\right]^2\right\}. 
\end{equation}
It is not difficult to show that the trial densities for the primitive FPI
and PA-FPI methods are identical (after normalization) but we shall employ
formula~(\ref{eq:B2}) for the RW-FPI technique too. Practice shows that
the penalty for considering the last two approximations is minimal,
while~(\ref{eq:B2}) has some advantages with regard to the organization
of the computations. 

To evaluate the partition functions for the primitive FPI approach, we
integrate~(\ref{eq:B1}) and obtain
\begin{equation}
\label{eq:B3}
Z^{2n}_{Pr}(\beta)=\frac{1}{\beta \hbar \omega} \left[ \prod_{k=1}^{2n}\frac{1}{\xi_k}\right] \left\{1-\sum_{k=1}^{2n} \frac{\beta m_0 \omega^2 \sigma_k^2}{\xi_k^2}\left(\frac{1-(-1)^k}{k\pi}\right)^2 \right\}^{-1/2}.
\end{equation}
We leave for the reader the simple task of showing that the $2n$-th order PA-FPI density matrix has the form
\begin{equation}
\label{eq:B4}
Z^{2n}_{PA}(\beta)=Z^{2n}_{Pr}(\beta)\exp\left[-\frac{\beta^2\hbar^2 \omega^2}{2\pi^2}\left(\frac{\pi^2}{6}-\sum_{k=1}^{2n}\frac{1}{k^2}\right)\right].
\end{equation}
The RW-FPI method's partition function is similar to the one
for the primitive FPI method and is given by
\begin{equation}
\label{eq:B5}
Z^{2n}_{RW}(\beta)=\frac{1}{\beta \hbar \omega} \left[ \prod_{k=1}^{2n}\frac{1}{\xi_{n,k}}\right] \left\{1-\sum_{k=1}^{2n} \frac{\beta m_0 \omega^2 \sigma_{n,k}^2}{\xi_{n,k}^2}\left(\frac{1-(-1)^k}{k\pi}\right)^2 \right\}^{-1/2},
\end{equation}
where 
\[\xi_{n,k}^2=1+\beta m_0 \omega^2 \sigma^2_{n,k}/2.\]

\end{widetext}

\section{Metropolis sampling}
The Metropolis \emph{et al.}\cite{Met53, Kal86} sampling of a general
probability density $\rho(x)$ with $x \in \Omega$ consists of generating
a homogeneous Markov chain having the transition probability density 
\begin{eqnarray}
\label{eq:C1}
\tau(x'|x)=A(x'|x)T(x'|x)+\nonumber \\ \delta(x'-x)\int_{\Omega}[1-A(y|x)]T(y|x) \ud y,
\end{eqnarray}
where $T(x'|x)$ is a trial transition probability density which would generate an irreducible chain by itself, $\delta(x'-x)$ is the Dirac function, and the acceptance probability $A(x'|x)$ is given by the formula
\[
A(x'|x)=\min\left\{1, \frac{\rho(x')T(x|x')}{\rho(x)T(x'|x)}\right\}.
\]
This choice of $A(x'|x)$ is one of the many possible which satisfy the condition
\[A(x'|x)T(x'|x)\rho(x)=T(x|x')A(x|x')\rho(x').\]
The last relation implies that the Markov chain of transition probability density $\tau(x|x')$ satisfies the \emph{detailed balance} condition 
\[\tau(x'|x)\rho(x)=\tau(x|x')\rho(x'),\]
which by integration against~$x'$ and use of the normalization condition~$\int_{\Omega}\rho(x)\ud x =1$ shows that~$\rho(x)$ is a stationary distribution of the transition kernel $\tau(x'|x)$. Moreover, it can be shown that the associated Markov chain is ergodic and that this implies that~$\rho(x)$ is the unique stationary distribution.\cite{Fel50} Let us consider the stationary sequence $X_0, X_1, \ldots$ with $X_0$ having the distribution density $\rho(x)$ and $X_n$ having the conditional density $P(X_n=x'|X_{n-1}=x)=\tau(x|x')$. One can generate a sample $x_0, x_1, \ldots$ starting with any point $x_0$, by the Metropolis algorithm:
\begin{enumerate}
\item{ given $x_n$, generate $x_{n+1}$ from the probability density $T(x|x_n)$; }
\item{ compute $A(x_{n+1}|x_n)$;}
\item{ generate a random number $q$ uniformly on $[0,1]$;}
\item{ if $q\leq A(x_{n+1}|x_n)$, accept the move; otherwise, reject it.}  
\end{enumerate}
For the expected value $\mathbb{E}(f)=\int_{\Omega}\rho(x)f(x)\ud x$, Birkhoff's ergodic theorem (Theorem~2.1, Chapter~6 of Ref.~\onlinecite{Dur96}) guarantees that
\begin{equation}
\label{eq:C2}
 \frac{1}{n}\sum_{k=0}^{n-1}f(X_i) \rightarrow \mathbb{E}(f)
\end{equation}
almost surely.  In words, the probability that we may generate a sequence $x_0, x_1, \ldots$ by the Metropolis algorithm such that 
\[ \frac{1}{n}\sum_{k=0}^{n-1}f(x_i) \nrightarrow \mathbb{E}(f)\] is zero. 
In fact, if the variance of $f(x)$ is finite 
\[\sigma_0^2(f)=\mathbb{E}(f-\mathbb{E}f)^2 < \infty,\]
a central limit theorem holds. Since the random variables $f(X_0)$ and $f(X_n)$ have the same distribution, their correlation coefficient takes the form
\[r_n(f)=\frac{\mathbb{E}\left[f(X_0)f(X_n)\right]-\mathbb{E}(f)^2}{\sigma_0^2(f)}\]
Explicitly, let us introduce the notation
\[\tau^{n}(x'|x)=\int_{\Omega}\ud x_1 \ldots \int_{\Omega}\ud x_{n-1} \tau(x'|x_1)\ldots \tau(x_{n-1}|x),\] 
with $\tau^0(x'|x)=\delta(x'-x)$ and $\tau^1(x'|x)=\tau(x'|x)$. 
Then, \[{\mathbb{E}\left[f(X_0)f(X_n)\right]=\int_{\Omega}\ud x \int_{\Omega}\ud x' \rho(x)\,\tau^n(x'|x)f(x)f(x')}.\]
In practice, we can evaluate these expectations, and therefore the correlation coefficients, again with the help of Birkhoff's theorem:
\begin{equation}
\label{eq:C3}
{\mathbb{E}\left[f(X_0)f(X_n)\right]=\lim_{k \to \infty} \frac{1}{k} \sum_{j=0}^{k-1} f(x_j)f(x_{j+n})}.
\end{equation}

In these conditions, it can be shown (Theorem~7.6, Chapter~7 of Ref.~\onlinecite{Dur96}) that
\begin{equation}
\label{eq:C4}
 \frac{\sum_{k=0}^{n-1}f(X_i)- \mathbb{E}(f)}{\sigma(f)\, n^{1/2}}\Rightarrow \xi,
\end{equation}
where $\xi$ has the standard normal distribution and 
\begin{equation}
\label{eq:C5}
\sigma^2(f)= \sigma_0^2(f)\left[1+2\sum_{n=1}^\infty r_n(f)\right].
\end{equation}
If the sampling were independent, the correlation coefficients would vanish and we would recover the classical central limit theorem. In practice however, the correlation coefficients are positive, many times having a slow decay to zero and the independent sampling may be considered a fortunate case. Without entering the details, we mention that there are two factors that contribute to large correlation coefficients: a) a strongly correlated proposal $T(x'|x)$ and b) a low overall efficiency. The overall efficiency (or the acceptance ratio) is defined as 
\begin{equation}
\label{eq:C6}
Ac=\int_{\Omega}\ud x' \!\! \int_{\Omega} \ud x \rho(x) A(x'|x) T(x'|x)
\end{equation}
and represents the fraction of moves accepted. Therefore, if the overall efficiency has large enough values ($Ac \geq 0.2$), it is a good idea to use an independent proposal from a trial probability $\rho_{tr}(x)$. If $\rho_{tr}(x)\approx \rho(x)$ and $f(x)$ is smooth enough, we may approximately relate the correlation coefficients to the overall efficiency as follows: 
from the relation~(\ref{eq:C1}), we easily compute
\begin{widetext}
\begin{eqnarray*}
r_1(f)=1- \frac{\int_{\Omega}\ud x' \!\! \int_{\Omega} \ud x [f(x)^2-f(x)f(x')] \rho(x')\rho_{tr}(x) A(x'|x)}{\int_{\Omega}\ud x' \!\! \int_{\Omega} \ud x [f(x)^2-f(x)f(x')] \rho(x')\rho(x) },
\end{eqnarray*}
\end{widetext}
where
\begin{equation}
\label{eq:C7}
A(x'|x)=\min\left\{1, \frac{\rho(x')\rho_{tr}(x)}{\rho(x)\rho_{tr}(x')}\right\}.
\end{equation}
Using the approximation $\rho(x')\rho_{tr}(x)A(x'|x)\approx Ac \, \rho(x)\rho(x')$, the right-hand side simplifies to $r_1(f)\approx 1-Ac$. In general, by a similar line of thought, one may argue that $r_n(f)\approx (1-Ac)^n $. The formula~(\ref{eq:C5}) takes the approximate value
\begin{equation}
\label{eq:C8}
 \sigma^2(f)\approx \sigma^2_0(f)\left[1+2\sum_{k=1}^\infty (1-Ac)^k\right] = \sigma^2_0(f)\left(\frac{2}{Ac}-1\right).
 \end{equation} 
Therefore, the bigger the acceptance probability, the faster the convergence of the Monte Carlo procedure. In the limit $Ac=1$, we recover the independent sampling, but a quick look at formula~(\ref{eq:C7}) shows that in this case $\rho_{tr}(x)=\rho(x)$.  

\section{Computation of the path weights $\sigma_{n,k}^2$ for the RW-FPI method.}

If $n<k\leq 2n$, we have
\begin{equation}
\label{eq:D1}
\sigma_{n,k}^2=\frac{2\beta\hbar^2}{\pi^2m_0} \sum_{j=0}^\infty
\frac{1}{(k+jn)^2}=\frac{2\beta\hbar^2}{\pi^2m_0}\frac{1}{n^2} \,
\text{h}\! \left(\frac{k-n}{n}\right), 
\end{equation}
where 
\[\text{h}(x)=\sum_{j=1}^\infty \frac{1}{(j+x)^2}.\]
Clearly, the values of the function~$\text{h}(x)$ are only needed over the interval $[0,1]$ and they can be evaluated via the Hurwitz $\zeta$-function, usually implemented by many mathematical libraries. Alternatively, $\text{h}(x)$ can be evaluated via the trivial identity
\begin{equation}
\label{eq:D2}
\text{h}(x)=\zeta(2)-2x \zeta(3)+ 3 x^2 \zeta(4) - x^3 \sum_{j=1}^\infty \frac{4j+3x}{(j+x)^2j^4}, 
\end{equation}
where $\zeta(s)$ is the Riemann $\zeta$-function
\[
\zeta(s)=\sum_{n=1}^\infty \frac{1}{n^s}.
\]
We have $\zeta(2)=\pi^2/6$, $\zeta(3)\approx 1.2020569031596$, and $\zeta(4)=\pi^4/90$, with the last series in~(\ref{eq:D2}) converging quite fast. More precisely, the error in the evaluation of $\text{h}(x)$ committed by truncating the series to the first~$n$ terms is easily seen to be smaller than $\sum_{j > n}4/j^5 \leq 1/n^4 $ uniformly on the whole interval $[0,1]$, so that summation over the first $100$ terms gives the value of $\text{h}(x)$ with an error of at most $10^{-8}$. This error is sufficiently small for our applications. 

\section{A specialized Monte Carlo scheme}

	As suggested in Appendix~C, the use of an independent trial 
distribution in the 
Metropolis algorithm is a good strategy provided that we are 
able to find a good approximation $\rho_{tr}^{4n+2}(x, \bar{a};\beta)$ to
the density we need to sample in this case,
\begin{equation}
\label{eq:38}
\rho^{4n+2}_{Mt}(x, \bar{a};\beta)= \frac{X_{Mt}^{4n+2}(x, \bar{a}, \beta)}{(2\pi)^{2n+1}}\exp\left(-\frac{1}{2}\sum_{k=1}^{4n+2} a_k^2\right).
\end{equation} 
This approximation may be taken to be the similar expression 
for a harmonic oscillator potential $m_0\omega^2(x-A)^2/2$, because we
know how to generate an independent sample of this. In order for the
approximation to work well for many of the single well potentials of
interest, we optimize the parameters $\omega$ and $A$ to obtain a best
fit in the sense of increasing the overall acceptance ratio. However,
since we are analyzing groundstate problems, sufficiently good
approximations can be obtained from the Ritz variational principle. Thus,
we look for the  parameters~$\omega$ and~$A$ which realize the minimum of
the functional
\[ E(\omega, A) = \int_{\mathbb{R}}\psi_{\omega, A} \hat{H}\psi_{\omega,A} \ud x,\] where \[ \psi_{\omega,A} (x)= \left(\frac{m_0\omega}{\pi \hbar}\right)^{1/4}\exp\left[-\frac{m_0\omega}{2\hbar}(x-A)^2\right] \] is the groundstate eigenfunction of the trial harmonic potential and
\[\hat{H} = -\frac{\hbar^2}{2m_0}\Delta + V(x)\] is the Hamiltonian of 
the original single well potential. By a translation of the
reference system, we may assume that the optimizing parameter~$A$ is
zero. For the case of the quartic potential $V(x)= x^4/2$, the best
optimizing parameters are $\omega = 1.442$ and $A=0$. Fig.~6 plots the
quartic potential and its best quadratic approximation. 
\begin{figure}[!tbp] 
   \includegraphics[clip=t]{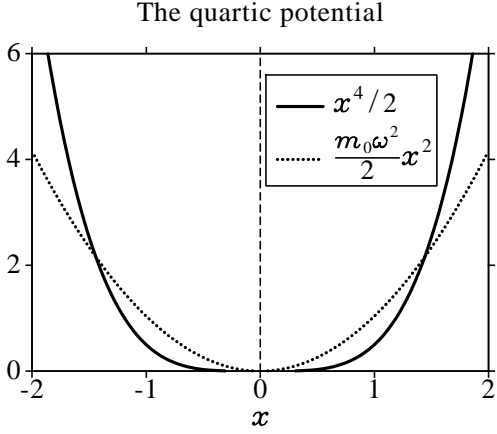} 
 \caption[sqr]
{\label{Fig:6}
A plot of the quartic potential (solid line) and its best variational quadratic approximation. Here, $m_0=1$ and $\omega=1.442$.
}
\end{figure}

Rather than using the $4n+2$-th order probability density of the best harmonic 
reference as the trial density, it is more convenient to use the slightly 
modified formula~(\ref{eq:B2}) of Appendix~B. The advantage is that~(\ref{eq:B2}) 
is the exponential of a series. As such, if we generate the 
vector $(x, a_0, ..., a_{4n+2})$ from the probability 
density $\rho_{tr}^{4n+2}(x,\bar{a};\beta)$ given by~(\ref{eq:B2}), 
we can use the vectors of the form $(x, a_0, ... , a_{4k+2})$ with $k\leq n$ 
for the paths of smaller length because it is clear that these vectors are drawn 
from the distribution $\rho_{tr}^{4k+2}(x,\bar{a};\beta)$. The time saved 
with the generation of random numbers fully compensates the slight decrease in 
the acceptance ratio.  We use~(\ref{eq:B2}) for all FPI methods in the following examples.

For PA-FPI and primitive FPI, there is another advantage in 
using the trial density~(\ref{eq:B2}). A large portion of the 
computational 
time is spent with the construction of the paths 
\[ \bar{a}_{4n+2}(u;\beta)=\sum_{k=1}^{4n+2}a_k\sigma_k \sin(k \pi u),\] 
especially for large~$n$. However, if the trial probability
density~(\ref{eq:B2}) is used, we can employ the recurrence formula
\[ \bar{a}_{4n+2}(u;\beta)=\bar{a}_{4n-2}(u;\beta)+\sum_{k=4n-1}^{4n+2}a_k\sigma_k \sin(k \pi u).\] Therefore, the time necessary to construct all the paths of length~$4k+2$ with $1
\leq k \leq n$ at a given point~$t$ scales like $\mathcal{O}(n)$ 
instead of $\mathcal{O}(n^2)$. This is especially important for the
PA-FPI method, which has the fastest convergence and for which a large
number of Monte Carlo steps is necessary to establish the asymptotic
convergence rate. Unfortunately, since the paths for RW-FPI are not
series, we cannot employ the same strategy for this method.

As shown in the Appendix~C, the advantage of our Monte Carlo 
strategy consists of the fact that it has low correlation provided that
the acceptance ratio is large. To a first approximation, the statistical
error in the estimation of the energy is [we employ the usual $2\sigma$
definition for the error, corresponding to a confidence interval of
$95.4\%$]
\begin{equation}
\label{eq:39}
\text{Err}_s\left(E^{4n+2}_{Mt}\right)=\frac{2\sigma_{0}\!\left(E^{4n+2}_{Mt}\right)}{\sqrt{N}}\left(\frac{2}{Ac}-1\right)^{1/2},
\end{equation}
where~$N$ is the number of Monte Carlo points, $\sigma_{0}^2\!\left(E^{4n+2}_{Mt}\right)$ is the variance of the T-estimator function, and~$Ac$ is the acceptance ratio [see~(\ref{eq:C8})]. A more precise formula is given by~(\ref{eq:C5}):
\begin{equation}
\label{eq:40}
\text{Err}_s\left(E^{4n+2}_{Mt}\right)=\frac{2\sigma_{0}\!\left(E^{4n+2}_{Mt}\right)}{\sqrt{N}}\left[1+2\sum_{k=1}^{\infty}r_{k}\left(E^{4n+2}_{Mt}\right)\right]^{1/2}
\end{equation} 
and we have shown in Appendix~C how the correlation coefficients can be evaluated during the Monte Carlo procedure. However, we can use~(\ref{eq:39}) to find the number of steps after which the correlation becomes negligible. In the case of the quartic potential, the acceptance ratio was bigger than~$0.6$ for all simulations performed. Since $2/0.6-1=2.333\approx 1+2\sum_{k=1}^{8}0.4^k=2.332$, we may safely truncate the series in~(\ref{eq:40}) to the first eight correlation coefficients and we shall do so for all computations concerning the quartic potential.  

Another important aspect in our computations is the numerical evaluation of 
the one-dimensional time averages that are involved. This issue was extensively studied by 
Sabo \emph{et. al.},\cite{Sab00} who concluded that a Gauss-Legendre quadrature 
in a number of points equal to three times the number of Fourier coefficients should 
suffice for most applications. We also employ the Gauss-Legendre quadrature scheme, 
but in a number of points equal to four times the \emph{maximum} number of Fourier 
coefficients computed. Extensive computer observations show that the relative error 
in the evaluation of the T-estimator function is smaller than $10^{-8}$ for the quartic 
potential. Of course, for real-life applications we do not need such a precision 
but here it is important to rule out any factor likely to alter the asymptotic law of 
convergence.

Earlier in this section, we saw that the scaling of the number of Monte Carlo points with the number of Fourier coefficients was dictated by the decay of the differences $E^{4n-2}_{Mt}-E^{4n+2}_{Mt}$, which we shall denote by~$DE^{4n+2}_{Mt}$. We shall improve on this fact by directly evaluating these differences with the help of a biased estimator. Define
\begin{widetext}
\begin{equation}
\label{eq:41}
r_{Mt}^{4n+2}(x,\bar{a};\beta)= X_{Mt}^{4n-2}(x,\bar{a};\beta)/X_{Mt}^{4n+2}(x,\bar{a};\beta)
\end{equation}
and
\begin{equation}
\label{eq:42}
R_{Mt}^{4n+2} =\frac{\int_{\mathbb{R}}\ud x\int_{\Omega}\ud P[\bar{a}] X_{4n+2}(x,\bar{a};\beta)r_{Mt}^{4n+2}(x,\bar{a};\beta) }{\int_{\mathbb{R}}\ud x\int_{\Omega}\ud P[\bar{a}] X_{4n+2}(x, \bar{a};\beta)}.
\end{equation}
Next, define 
\begin{equation}
\label{eq:43}
DE^{T,Mt}_{4n+2}(x,\bar{a};\beta)=E^{T,Mt}_{4n-2}(x,\bar{a};\beta)\,r_{Mt}^{4n+2}(x,\bar{a};\beta)/R_{Mt}^{4n+2} -E^{T,Mt}_{4n+2}(x,\bar{a};\beta).
\end{equation}
It is a simple exercise to show that
\begin{equation}
\label{eq:44}
E^{4n-2}_{Mt}-E^{4n+2}_{Mt} =\frac{\int_{\mathbb{R}}\ud x\int_{\Omega}\ud P[\bar{a}] X_{4n+2}(x,\bar{a};\beta)DE_{4n+2}^{T,Mt}(x,\bar{a};\beta) }{\int_{\mathbb{R}}\ud x\int_{\Omega}\ud P[\bar{a}] X_{4n+2}(x, \bar{a};\beta)}.
\end{equation}

A biased estimator for the function~(\ref{eq:43}) can be constructed as follows: assume you are given a sequence $(x_k,\bar{a}_k)$ with $1 \leq k \leq N$, which samples the probability distribution~(\ref{eq:38}). At step~$k$, compute 
\[ 
R_{Mt}^{k,4n+2}=\frac{1}{k}\sum_{j=1}^{k}r_{Mt}^{4n+2}(x_j,\bar{a}_j;\beta)\quad \text{and} \quad \text{Err}_s(k, R_{Mt}^{4n+2}).
\]
and construct the function
\begin{equation}
\label{eq:45}
DE^{T,Mt}_{k,4n+2}(x,\bar{a};\beta)=E^{T,Mt}_{4n-2}(x,\bar{a};\beta)\,r_{Mt}^{4n+2}(x,\bar{a};\beta)/R_{Mt}^{k,4n+2} -E^{T,Mt}_{4n+2}(x,\bar{a};\beta).
\end{equation}  
Then, the biased estimator is defined by the well-known recurrence formula 
\begin{equation}
\label{eq:46}
DE^{k,4n+2}_{Mt}=\left[(k-1)DE^{k-1,4n+2}_{Mt}+DE^{T,Mt}_{k,4n+2}(x_{k}, \bar{a}_{k};\beta)\right]\Big/k
\end{equation}
starting with $DE^{0,4n+2}_{Mt}=0$. Clearly, $DE^{k,4n+2}_{Mt}$ converges to $DE^{4n+2}_{Mt}$ as~$k$ gets large.

  The bias in~(\ref{eq:45}) is due to the fact that we do not use the exact value of~$R_{Mt}^{4n+2}$ but its unbiased statistical estimator. However, for large enough~$k$, it is not difficult to justify the estimate:
\[
\left|DE^{T,Mt}_{k,4n+2}(x,\bar{a};\beta)-DE^{T,Mt}_{4n+2}(x,\bar{a};\beta)\right|\lessapprox \frac{\big|E^{T,Mt}_{4n-2}(x,\bar{a};\beta)\big|\,r_{Mt}^{4n+2}(x,\bar{a};\beta)}{R_{Mt}^{4n+2}}\frac{\text{Err}_s(k,R_{Mt}^{4n+2})}{R_{Mt}^{4n+2}}.
\] 
It follows then that the error due to bias is at most
\begin{equation}
\label{eq:47}
\text{Err}_b(N, DE^{4n+2}_{Mt})= \frac{1}{N}\sum_{k=1}^N\frac{\big|E^{T,Mt}_{4n-2}(x_k,\bar{a}_k;\beta)\big|\,r_{Mt}^{4n+2}(x_k,\bar{a}_k;\beta)}{R_{Mt}^{k,4n+2}}\frac{\text{Err}_s(k,R_{Mt}^{4n+2})}{R_{Mt}^{k,4n+2}}.
\end{equation}
The total error is then obtained by also adding the statistical error computed with the help of the formula~(\ref{eq:40}):
\begin{equation}
\label{eq:48}
\text{Err}(N, DE^{4n+2}_{Mt})=\text{Err}_s(N, DE^{4n+2}_{Mt})+\text{Err}_b(N, DE^{4n+2}_{Mt}).
\end{equation}

In the present paper, we pre-computed a start value of~$R_{Mt}^{4n+2}$ using a quarter of the number of Monte Carlo points during the warm-up step and then continued to improve the value in the main procedure. In these conditions, one may argue that the error for the difference~(\ref{eq:44}) satisfies the inequality
\begin{equation}
\label{eq:49}
\text{Err}(N, DE^{4n+2}_{Mt}) \leq \text{Err}_s(N, DE^{4n+2}_{Mt}) +  \sqrt{5}\,\mathbb{E}\left(\big|E^{4n-2}_{Mt}\big|\right) \frac{\text{Err}_s(5N/4,R_{Mt}^{4n+2})}{R_{Mt}^{4n+2}}, 
\end{equation}
where 
\begin{displaymath}
\mathbb{E}\left(\big|E^{4n-2}_{Mt}\big|\right)=\frac{\int_{\mathbb{R}}\ud x\int_{\Omega}\ud P[\bar{a}] X_{4n-2}(x,\bar{a};\beta)\,\big|E^{T,Mt}_{4n-2}(x,\bar{a};\beta)\big| }{\int_{\mathbb{R}}\ud x\int_{\Omega}\ud P[\bar{a}] X_{4n-2}(x, \bar{a};\beta)}.
\end{displaymath}
\end{widetext}
   
   Formula~(\ref{eq:49}) helps us explain why the use of the biased estimator~(\ref{eq:45}) is advantageous. Had we directly evaluated the difference
\begin{equation}
\label{eq:50}
DE^{4n+2}_{Mt}=E^{4n-2}_{Mt}-E^{4n+2}_{Mt},
\end{equation}
the error would have been
\begin{equation}
\label{eq:51}
\text{Err} \left(N,DE^{4n+2}_{Mt}\right)=\text{Err}_s \left(N,E^{4n-2}_{Mt}\right)+ \text{Err}_s \left(N,E^{4n+2}_{Mt}\right).
\end{equation}
 Notice however that both $r_{Mt}^{4n+2}(x,\bar{a}; \beta)$ and $DE_{4n+2}^{T,Mt}(x,\bar{a};\beta)$ converge to 1 and 0, respectively as $n \to \infty$. In turn, their variances (which control the statistical errors) converge to zero. Clearly, this is not the case for the variance of the T-method energy estimator. More precisely, Table~\ref{Tab:I} presents strong numerical evidence suggesting that the decay of their standard deviations is as fast as $\mathcal{O}(1/n^2)$ and we expect this to be true for all smooth enough potentials. This implies that for a fixed but large number of Monte Carlo points~$N$, the error in~(\ref{eq:48}) has the asymptotic behavior
\begin{equation}
\label{eq:52}
\text{Err}(N, DE_{4n+2}^{T,Mt})\approx \frac{const}{n^2\sqrt{N}}
\end{equation}
  
The importance of~(\ref{eq:52}) is twofold. First, it shows that if the estimator~(\ref{eq:46}) is used, the scaling of the number of Monte Carlo samples with respect to the number of Fourier coefficients is now determined by the decay of $E^{4n+2}_{Mt}-E$ to zero. More precisely, we have $N\propto n^6$ for PA-FPI and RW-FPI, $N\propto n^4$ for TT-FPI, and $N\propto n^2$ for
primitive FPI. 

Second, the errors of the estimators of order one and two [see~(\ref{eq:35}) and~(\ref{eq:36})] have the asymptotic behavior:
\begin{eqnarray}
\label{eq:53}
\text{Err} (N, FE_{4n+2}^{T,Mt})&=& \text{Err}_s (N, E_{4n+2}^{T,Mt})+  \frac{1}{\alpha}\nonumber  \frac{const}{n\sqrt{N}} \approx \\ &\approx &\text{Err} (N, E_{4n+2}^{T,Mt}) 
\end{eqnarray} 
and, 
\begin{eqnarray}
\label{eq:54}
\text{Err} (N, SE_{4n+2}^{T,Mt})=\text{Err}_s (N, E_{4n+2}^{T,Mt})+ \nonumber\\
\frac{const}{\alpha(\alpha+1)\sqrt{N}} \bigg[ \frac{2\alpha+1}{n} + \nonumber
\frac{1}{n-1}+ 2+\frac{2n^2}{(n-1)^2}\bigg] \approx \\ \approx \text{Err}_s (N, E_{4n+2}^{T,Mt})+ \frac{4\cdot const}{\alpha(\alpha+1)\sqrt{N}} \quad
\end{eqnarray} 
This readily implies that the use of the estimators of order one and 
two does \emph{not} change the scaling of the number of Monte Carlo points 
needed to achieve a given error threshold for the estimated energy with the number 
of Fourier coefficients. The net result is an improvement in the asymptotic behavior 
for the estimators of order one and two. However, in the case of the second-order 
estimator, we notice an increase in the variance of the estimator which may be quite 
large for practical purposes. For the first-order 
estimator \emph{there is no asymptotic increase in the variance}, which makes it 
more suitable for practical applications. In fact, the first-order estimator may also be used for potentials that do not have continuous second-order derivatives but for which the decay with the number of Fourier coefficients implied by~(\ref{eq:52}) can be replaced by the slower one
\begin{displaymath}
\text{Err}(N, DE_{4n+2}^{T,Mt})\approx \frac{const}{n\sqrt{N}}.
\end{displaymath}
Finally, in the cases where it cannot be utilized as an energy estimator because of an unduly large variance,  the correction term brought in by the first-order estimator is still useful as a measure of how far the zero-order estimator is from the true result. 

  The reader may work out the expression 
for the estimator of order three and see that in this case the scaling is changed. 
This explains our earlier assertion that the estimators of order three or more 
are of little practical value.

\section{Tables of numerical values}

The following tables contain the numerical results described in Section
IVB.  See that discussion for the details.

\begin{table*}[!bh]
\caption{ \label{Tab:II} Average energies, estimated differences, and their statistical error for the quartic potential at $\beta =10$. The variational energy is $0.530183$. }
\begin{tabular}{|c |c |c |c |c |c |c |c |c |c |c |c |c | }
\hline
$n$ & 1 & 2 & 3 & 4 & 5 & 6 & 7 & 8 & 9 & 10 & 11 & 12\\ 
\hline \hline
\multicolumn{13}{|c|}{\emph{Average energies}}\\
\hline
$E_{Pr}^{4n+2}$& 0.302878&0.365234&0.401528&0.425003&0.441342&0.453379&
               	 0.462581&0.469834&0.475704&0.480548&0.484613&0.488071\\
\hline
$E_{TT}^{4n+2}$& 0.343731&0.416263&0.454541&0.476808&0.490728&0.499978&
               	 0.506376&0.510972&0.514391&0.516994&0.518994&0.520602\\
\hline
$E_{RW}^{4n+2}$& 0.351676&0.432846&0.473011&0.493918&0.505667&0.512786&
               	 0.517363&0.520451&0.522627&0.524201&0.525370&0.526247\\
\hline
$E_{PA}^{4n+2}$& 0.596947&0.552843&0.541042&0.536268&0.533916&0.532629&
               	 0.531862&0.531383&0.531069&0.530854&0.530701&0.530593\\
\hline \hline
\multicolumn{13}{|c|}{\emph{Estimated differences}}\\
\hline
$DE_{Pr}^{4n+2}$& -.124981&-.062354&-.036316&-.023475&-.016353&-.012025&
               	  -.009205&-.007265&-.005872&-.004848&-.004062&-.003452\\
\hline
$DE_{TT}^{4n+2}$& -.147346&-.072843&-.038307&-.022241&-.013929&-.009222&
               	  -.006401&-.0046052&-.003424&-.002593&-.002013&-.001589\\
\hline
$DE_{RW}^{4n+2}$& -.162238&-.080976&-.040075&-.020896&-.011746&-.007111&
               	  -.004573&-.003100&-.002176&-.001575&-.001168&-.000886\\
\hline
$DE_{PA}^{4n+2}$& 0.549021&0.044095&0.011781&0.004772&0.002347&0.001285&
               	  0.000763&0.000481&0.000316&0.000216&0.000151&0.000109\\
\hline \hline
\multicolumn{13}{|c|}{\emph{Statistical errors for energies}($2\sigma$)}\\
\hline
$E_{Pr}^{4n+2}$& 0.000088&0.000084&0.000081&0.000080&0.000078&0.000078&
               	 0.000077&0.000077&0.000076&0.000076&0.000076&0.000076\\
\hline
$E_{TT}^{4n+2}$& 0.000056&0.000057&0.000056&0.000055&0.000054&0.000054&
               	 0.000054&0.000053&0.000053&0.000053&0.000053&0.000053\\
\hline
$E_{RW}^{4n+2}$& 0.000043&0.000042&0.000041&0.000040&0.000039&0.000038&
               	 0.000038&0.000038&0.000038&0.000037&0.000037&0.000037\\
\hline
$E_{PA}^{4n+2}$& 0.000024&0.000023&0.000021&0.000020&0.000020&0.000020&
               	 0.000020&0.000019&0.000019&0.000019&0.000018&0.000018\\
\hline \hline
\multicolumn{13}{|c|}{\emph{Statistical errors for differences }($2\sigma$)}\\
\hline
$DE_{Pr}^{4n+2}$& 0.032356&0.000304&0.000095&0.000055&0.000036&0.000026&
                  0.000020&0.000016&0.000012&0.000010&0.000009&0.000007\\
\hline
$DE_{TT}^{4n+2}$& 0.054577&0.001325&0.000199&0.000092&0.000061&0.000046&
                  0.000036&0.000030&0.000025&0.000021&0.000018&0.000015\\
\hline
$DE_{RW}^{4n+2}$& 0.037799&0.001600&0.000333&0.000074&0.000045&0.000032&
                  0.000025&0.000019&0.000016&0.000013&0.000011&0.000009\\
\hline
$DE_{PA}^{4n+2}$& 0.004595&0.000080&0.000027&0.000014&0.000009&0.000007&
               	  0.000005&0.000004&0.000003&0.000003&0.000002&0.000002\\
\hline
\end{tabular}
\end{table*}

\begin{table*}[!thb]
\caption{\label{Tab:I} Standard deviations for $r_{Mt}^{4n+2}(x,\bar{a};\beta)$ and $DE^{T,Mt}_{4n+1}(x,\bar{a};\beta)$ and their asymptotic convergence exponents $\alpha$.}
\begin{tabular}{|c |c |c |c |c |c |c |c |c |c |c |c |c | }
\hline
$n$ & 2 & 3 & 4 & 5 & 6 & 7 & 8 & 9 & 10 & 11 & 12 &$\alpha$\\ 
\hline \hline
\multicolumn{13}{|c|}{\emph{Primitive FPI}}\\
\hline
$r_{Pr}^{4n+2}$&1.771976&0.437892&0.226571&0.143267&0.099735&
               	 0.073744&0.056769&0.045052&0.036617&0.030341&0.025542&2.027\\
\hline
$DE^{T,Pr}_{4n+2}$&0.907676&0.222962&0.116896&0.075168&0.052794&
               	 0.039381&0.030502&0.024296&0.019827&0.016464&0.013907&1.985\\
\hline \hline
\multicolumn{13}{|c|}{\emph{TT-FPI}}\\
\hline
$r_{TT}^{4n+2}$&9.411364&1.149898&0.499554&0.326273&0.241282&
               	 0.187604&0.151356&0.125104&0.105345&0.090015&0.077867&1.822\\
\hline
$DE^{T,TT}_{4n+2}$&5.259389&0.856299&0.342638&0.204763&0.148354&
               	 0.112764&0.089384&0.072996&0.060898&0.051628&0.044358&1.825\\
\hline \hline
\multicolumn{13}{|c|}{\emph{RW-FPI}}\\
\hline
$r_{RW}^{4n+2}$&17.63636&1.953891&0.543819&0.324225&0.228415&
               	 0.171841&0.134647&0.108553&0.089442&0.074985&0.063776&1.961\\
\hline
$DE^{T,RW}_{4n+2}$&8.421340&2.378073&0.393884&0.210005&0.142382&
               	 0.105825&0.082139&0.065720&0.053826&0.044884&0.037983&1.999\\
\hline \hline
\multicolumn{13}{|c|}{\emph{PA-FPI }}\\
\hline
$r_{PA}^{4n+2}$&1.164360&0.309710&0.182490&0.122904&0.088704&
               	 0.067010&0.052355&0.041982&0.034380&0.028648&0.024221&2.100\\
\hline
$DE^{T,PA}_{4n+2}$&1.596240&0.291506&0.142120&0.087909&0.060274&
               	 0.044039&0.033599&0.026465&0.021387&0.017629&0.014785&2.013\\
\hline
\end{tabular}
\end{table*}

\end{document}